\documentclass[iop,revtex4]{emulateapj} \newcommand{\ltab}{\LongTables}

\input{colordvi}

\usepackage{hyperref,enumerate,numdef,color}
\hypersetup{
    unicode=false,                  
    pdftoolbar=true,                
    pdfmenubar=true,                
    pdffitwindow=true,              
    pdfstartview={FitH},            
    pdftitle={HE0020-1741},         
    pdfauthor={vplacco},            
    pdfsubject={Astronomy},         
    pdfcreator={dvipdf},            
    pdfproducer={dvipdf},           
    pdfkeywords={metal-poor stars}, 
    pdfnewwindow=true,              
    colorlinks=true,                
    linkcolor=red,                  
    citecolor=blue,                 
    filecolor=magenta,              
    urlcolor=cyan,                  
    breaklinks=true,
    linktocpage
}

\num\newcommand{\s01}{\object{HE~0020$-$1741}}
\num\newcommand{\s02}{\object{G77$-$61}}
\num\newcommand{\s03}{\object{BD$+$44$^\circ$493}}
\num\newcommand{\s04}{\object{HE~1327$-$2326}}
\num\newcommand{\s05}{\object{HE~2139$-$5432}}
\num\newcommand{\s06}{\object{CS~22949$-$037}}
\num\newcommand{\s07}{\object{CD$-$38$^\circ$245}}
\num\newcommand{\s08}{\object{CS~30336$-$049}}
\num\newcommand{\s09}{\object{HE~0107$-$5240}}
\num\newcommand{\s10}{\object{HE~0057$-$5959}}
\num\newcommand{\s11}{\object{HE~1310$-$0536}}
\num\newcommand{\s12}{\object{SDSS~J1313$-$0019}}

\newcommand{\eps}[1]{\ensuremath{\log\epsilon\,(\mathrm{#1})}}
\newcommand{\xx}{{\tablenotemark{a}}}
\newcommand{\yy}{{\tablenotemark{b}}}

\newcommand{\abund}[2]{\ensuremath{[\mathrm{#1}/\mathrm{#2}]}}
\newcommand{\cfe}{\abund{C}{Fe}}

\newcommand{\xfe}[1]{\abund{#1}{Fe}}

\newcommand{\metal}{\abund{Fe}{H}}

\newcommand{\teff}{\ensuremath{T_\mathrm{eff}}}
\newcommand{\logg}{\ensuremath{\log\,g}}
\newcommand{\Msun}{\mathrm{M}_\odot}

\newcommand{\erg}{\ensuremath{\mathrm{erg}}}

\newcommand{\bd}{\object{BD+44$^\circ$493}}
\newcommand{\hes}{\object{HE~0020$-$1741}}

\newcommand{\hkl}{\object{BPS~CS~30324$-$0063}}
\newcommand{\ravel}{\object{RAVE~J002244.9$-$172429}}
\newcommand{\twoml}{\object{2MASS~J00224486$-$1724290}}

\shorttitle{Observational Constraints on First-Star Nucleosynthesis. II.}
\shortauthors{Placco et al.}

\begin{document}

\title{
Observational Constraints on First-Star Nucleosynthesis. II.\\
Spectroscopy of an Ultra Metal-Poor CEMP-no Star\footnotemark[1]
}


\footnotetext[1]{Based on observations gathered with: The 6.5 meter Magellan
Telescopes located at Las Campanas Observatory, Chile; and the New Technology
Telescope (NTT) of the European Southern Observatory (088.D-0344A), La Silla,
Chile.}

\author{
Vinicius M.\ Placco\altaffilmark{2,3},
Anna         Frebel\altaffilmark{3,4},
Timothy   C.\ Beers\altaffilmark{2,3},
Jinmi          Yoon\altaffilmark{2,3},\\
Anirudh       Chiti\altaffilmark{3,4},
Alexander     Heger\altaffilmark{5,6,7}
Conrad         Chan\altaffilmark{5},\\
Andrew    R.\ Casey\altaffilmark{8},
Norbert  Christlieb\altaffilmark{9}}

\altaffiltext{2}{Department of Physics, University of Notre Dame, 
                 Notre Dame, IN 46556, USA}
\altaffiltext{3}{JINA Center for the Evolution of the Elements, USA}
\altaffiltext{4}{Department of Physics and Kavli Institute for Astrophysics
                 and Space Research, Massachusetts Institute of
                 Technology, Cambridge, MA 02139, USA}
\altaffiltext{5}{Monash Centre for Astrophysics, School of Physics and Astronomy,
                 19 Rainforest Walk, Monash University, Vic 3800, Australia}
\altaffiltext{6}{Shanghai Jiao-Tong University, CNA, Department of
                 Physics and Astronomy, Shanghai 200240, P.~R.~China}
\altaffiltext{7}{University of Minnesota, School of Physics and Astronomy,
                 Minneapolis, MN 55455, USA}
\altaffiltext{8}{Institute of Astronomy, University of Cambridge, Madingley
				 Road, Cambridge CB3 0HA, UK}
\altaffiltext{9}{Zentrum f\"ur Astronomie der Universit\"at Heidelberg, 
                 Landessternwarte, K\"onigstuhl 12, 69117, Heidelberg, Germany}

\addtocounter{footnote}{9}

\begin{abstract}

We report on the first high-resolution spectroscopic analysis of \hes, a bright
($V=12.9$), ultra metal-poor (\metal~=~$-$4.1), carbon-enhanced (\cfe~=~$+$1.7)
star selected from the Hamburg/ESO Survey. This star exhibits low abundances of
neutron-capture elements (\xfe{Ba}~=~$-1.1$), and an absolute carbon abundance
$A$(C) = 6.1; based on either criterion, \hes\ is sub-classified as a CEMP-no
star. We show that the light-element abundance pattern of \hes\ is consistent
with predicted yields from a massive (M~=~$21.5\,\Msun$), primordial composition,
supernova (SN) progenitor.  We also compare the abundance patterns of
other ultra metal-poor stars from the literature with available measures of C,
N, Na, Mg, and Fe abundances with an extensive grid of SN models
(covering the mass range $10\,\Msun-100\,\Msun$), in order to probe the nature
of their likely stellar progenitors. Our results suggest that at least two
classes of progenitors are required at \metal~$<-4.0$, as the abundance
patterns for more than half of the sample studied in this work (7 out of 12
stars) cannot be easily reproduced by the predicted yields.

\end{abstract}


\section{Introduction}
\label{intro}

Observational evidence has emerged over the past few decades indicating
that carbon is ubiquitous in the early Universe. The class of
carbon-enhanced metal-poor \citep[CEMP; \cfe$\geq+0.7$, e.g.,
][]{beers2005,aoki2007} stars are found with increasing fractions at
lower metallicities, and account for at least 80\% of all ultra
metal-poor (UMP; [Fe/H]\footnote{\abund{A}{B} = $log(N_X/{}N_Y)_{\star} -
\log(N_X/{}N_Y) _{\odot}$, where $N$ is the number density of atoms of
elements $X$ and $Y$ in the star ($\star$) and the Sun ($\odot$),
respectively.} $< -4.0$) stars observed to date \citep{lee2013,
placco2014c}. In particular, the so-called CEMP-no stars (which exhibit
sub-Solar abundances of neutron-capture elements; e.g., \xfe{Ba} $<$
0.0) are believed to be direct descendants from the very first stellar
generations formed after the Big Bang \citep{ito2013,spite2013,placco2014b,
hansen2016}. In addition, the discovery of high-redshift carbon-enhanced
damped Ly$\alpha$ systems \citep{cooke2011,cooke2012}, which present
qualitatively similar light-element (from C to Si) abundance patterns as
the CEMP-no stars, provides additional evidence that carbon is an
important contributor to the earliest chemical evolution.

One of the main open questions is whether the presence of carbon is
required for the formation of low-mass second-generation stars
\citep{frebel2007}. This idea can be tested with low-mass long-lived
UMP stars that are thought to have formed from an ISM polluted by the
nucleosynthesis products of massive metal-free Population III (Pop III)
stars. These massive stars could have formed as early as several hundred
million years after the Big Bang, at redshift $z \approx 20$
\citep{alvarez2006}. The recently discovered quasar ULAS~J1120$+$0641 at
redshift $z = 7.085$ \citep{simcoe2012}, with an overall metal
abundance (defined as elements heavier then helium) of
\abund{Z}{H}$\leq-4.0$, could be an example of a viable site for the
formation of the first stars. The lack of heavy metals may prevent the
formation of low-mass stars \citep[due to inefficient cooling;
][]{bromm2001}, supporting the suggestion that the early-Universe
initial mass function was strongly biased toward high-mass stars. In
the picture that has developed, it is these high-mass stars that would
quickly evolve and enrich the primordial ISM with elements heavier than
helium, including carbon \citep{meynet2006}.

Even though the relevance of CEMP-no stars as probes of the first stellar
generations in the Universe is well-established, the exact conditions that led
to their formation remain an active area of inquiry. There have been a number
of advances in the theoretical description of the likely stellar progenitors of
CEMP-no stars over the last decade \citep[see][for a recent review on the
subject]{nomoto2013}.
The suggested scenarios include 
the ``spinstar'' model \citep[rapidly-rotating, near zero-metallicity, massive
stars;][]{meynet2010,chiappini2013}, the ``faint SNe'' that undergo mixing and
fallback \citep{umeda2005,nomoto2006,tominaga2014}, and the metal-free massive
stars from \citet{heger2010}. 

In the spinstar model, it is assumed that the chemical composition of the
observed UMP stars is a combination of the evolution of the massive star itself
mixed with some amount of interstellar material \citep{meynet2006,meynet2010}.
It follows that the source of heavy metals in the UMP stars could arise
from a different set of progenitors. 
For the faint SNe and metal-free massive stars, the initial chemical abundances
of the progenitor mimic the primordial Big Bang Nucleosynthesis composition:
76\% hydrogen, about 24\% helium, and a trace amount of lithium.  In both cases,
the observed chemical elements in UMP stars were formed during the progenitor
stellar evolution, either by internal burning and/or explosive events, and their
abundance is the result of mixing between the SNe ejecta with surrounding
primordial gas.
These two models differ in terms of the treatment of the mixing and fallback of
processed materials in the progenitor, which varies with mass and explosion
energy \citep[see][for details]{tominaga2007}. The \citet{heger2010}
models compute explosion energy and fallback self-consistently based on a
hydrodynamic model, considering that the SN explosion is spherical.

All of the aforementioned models are able to reproduce a subset of (but
not all) of the observed elemental-abundance patterns of UMP stars
reasonably well. Nevertheless, the question of whether one or more
classes of progenitors were present (and their relative frequencies) in
the primordial Universe is still under discussion. In Paper~I of this series,
\citet{yoon2016} present evidence based on the morphology of the relationship
between the absolute abundance of carbon, $A$(C) $= \log\, \epsilon $(C), and
[Fe/H], coupled with clear differences in the absolute abundances of the light
elements Na and Mg among CEMP-no stars, that at least two classes of
progenitors are likely to be required. It appears that one class
(spinstars) dominates at the very lowest metallicities, [Fe/H] $
< -4.5$, whereas the other (faint SNe) dominates over the range
$-4.5 \le $ [Fe/H] $\le -2.5$. In the metallicity range $-5.0 \le$ [Fe/H] $\le
-4.0$, there are examples of stars that are associated with either.

For the above reasons, we suggest that the very best stars to place
constraints on the nature of the CEMP-no progenitors are the UMP stars
with metallicities between $-5.0$ and $-4.0$. Unfortunately, such stars
are still exceedingly rare \citep{yong2013b}. Even though their numbers
have increased considerably over the last decade \citep[21 stars
according to][]{placco2015}, many more are needed, in order to fully
understand the nature of their stellar progenitors and the associated
nucleosynthesis processes.

In this paper, we report on a high-resolution spectroscopic abundance analysis
\hes, a relatively bright ($V=12.89$) CEMP-no (\metal~$ = -4.05$,
\cfe~=~$+$1.74, and \xfe{Ba}~=~$-1.11$) star, which was first identified as the
metal-poor candidate CS~30324-0063 in the HK Survey
\citep{beers1985,beers1992}, and later re-identified in the Hamburg/ESO Survey
\citep[HES; ][]{christlieb2008} and the Radial Velocity Experiment \citep[RAVE;
][]{fulbright2010}. \hes\ was also studied by \citet{hansen2016}, focusing on
long-term radial-velocity monitoring of CEMP-no stars. We use the newly derived
abundance pattern of \hes, along with literature data for 11 other UMP and
hyper metal-poor (HMP; [Fe/H] $< -5.0$) stars, to assess evidence in support of
the conclusion in Paper~I that CEMP-no stars require more than one class of
stellar progenitors.

This paper is outlined as follows: Section~\ref{secobs} describes the
medium-resolution spectroscopic target selection and high-resolution
follow-up observations, followed by the determinations of the stellar
parameters and chemical abundances in Section~\ref{secatm}.
Section~\ref{disc} details our analysis of UMP and HMP stars from the
literature with a grid of supernova yields, and evaluates the impact of
these data on current hypotheses for chemical evolution in the early
Universe. Our conclusions are provided in Section~\ref{final}.

\section{Observations}
\label{secobs}

\begin{figure*}[!ht]
\epsscale{1.15}
\plotone{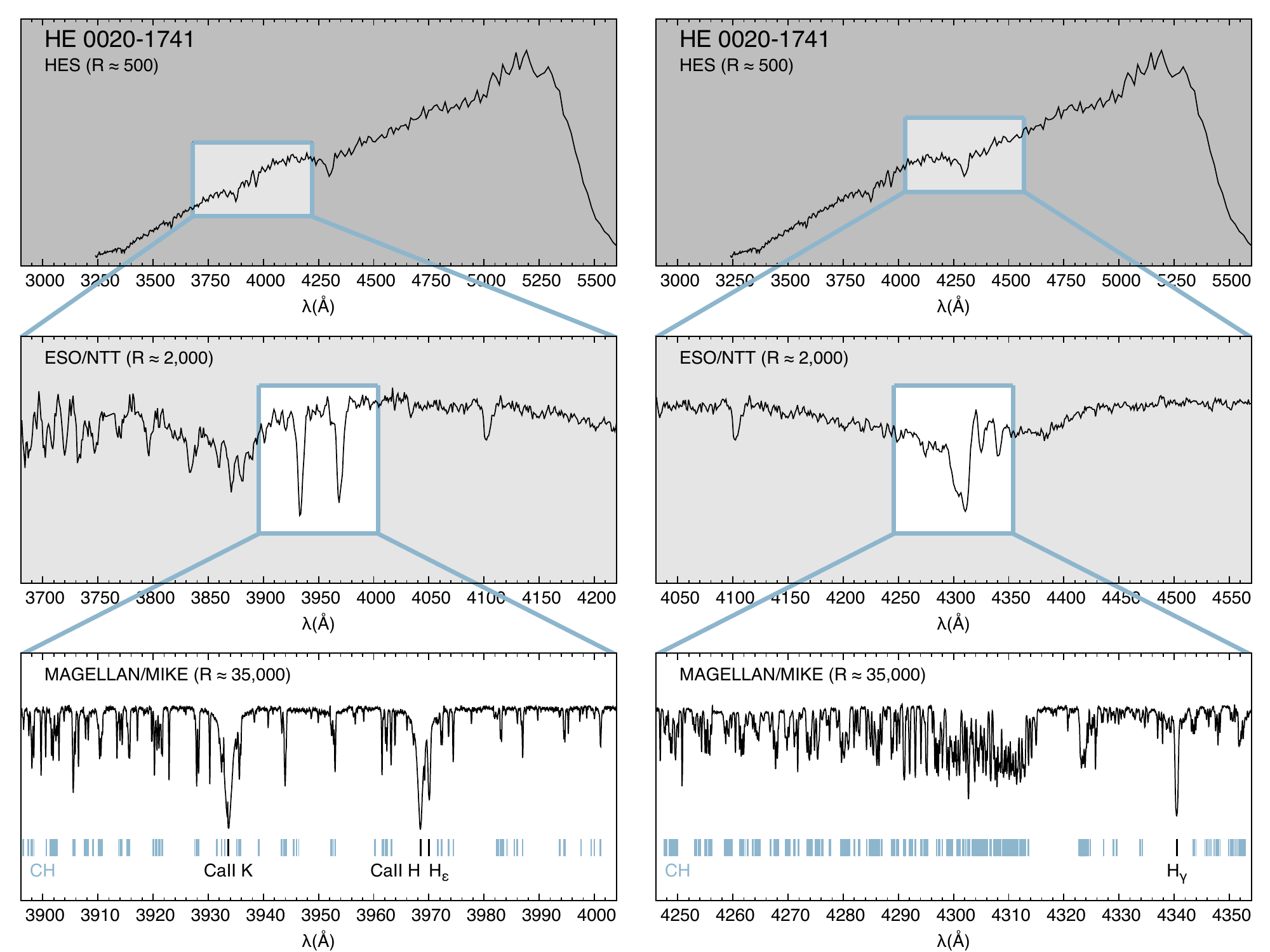}
\caption{Spectra of \protect\hes\ with three different resolving powers.
Upper panels: Low-resolution ($R\sim 500$) HES objective-prism spectrum. Middle panels:
Medium-resolution ($R\sim 2,000$) ESO/NTT spectrum. Lower panels:
High-resolution ($R\sim 35,000$) Magellan/MIKE spectrum. The left panels show a
zoom-in of the region including the Ca\,{\sc ii} H and K lines, and the right panels
show a zoom-in of the region surrounding the CH $G$-band. Interesting
strong features are identified in the lower panels; the blue lines
correspond to individual lines associated with the CH molecule.}
\label{highres}
\end{figure*}

The star \hes\ was selected as a metal-poor candidate by
\citet{christlieb2008}, based on its weak Ca\,{\sc ii}~K feature (3933\,
{\AA}, used as the primary metallicity indicator) in the HES objective-prism
spectrum. This star was also selected by \citet{placco2010}, based on
its strong CH $G$-band (4300\, {\AA}, the primary carbon-abundance
indicator). Figure~\ref{highres} is a comparison between the
low-resolution HES objective-prism spectrum ($R\sim 500$, upper panels),
the medium-resolution ($R\sim 2,000$, middle panels) spectrum, and the
high-resolution ($R\sim 35,000$, lower panels) spectrum of \hes. The
left panels show a zoom-in of the region surrounding the Ca\,{\sc ii}~K
line, and the right panels show a zoom-in of the region near the CH
$G$-band. Note the lack of measurable metallic features (other than Ca)
in the HES spectra, which is an indication of the low metallicity of the
target. The Balmer lines of hydrogen are also quite weak, suggesting
that the target has a cool effective temperature. However, as can be
seen from inspection of the medium- and high-resolution spectra, the
Ca\,{\sc ii} lines, as well as hydrogen lines from the Balmer series,
are clearly identifiable, as labeled in Figure~\ref{highres}. In
addition, the lower panels show a number of CH features in both regions,
which are used below to determine the carbon abundance from the
high-resolution spectrum.

\begin{deluxetable}{lr}[!ht]
\tablewidth{0pt}
\tabletypesize{\scriptsize}
\tablecaption{Observational Data \label{candlist}}
\tablehead{
\multicolumn{2}{c}{Identifiers}}
\startdata
&\\
HK Survey          & \hkl   \\
Hamburg/ESO Survey & \hes   \\
2MASS              & \twoml \\
RAVE               & \ravel \\
&\\
\hline
\multicolumn{2}{c}{Coordinates and Photometry} \\
\hline
&\\
$\alpha$ (J2000)  &    00:22:44.86  \\
$\delta$ (J2000)  & $-$17:24:29.07  \\
$V$ (mag)         &          12.89  \\
$B-V$             &           0.94  \\
&\\
\hline
\multicolumn{2}{c}{RAVE} \\
\hline
&\\
$R$               &    $\sim$8,000  \\
v$_{r}$(km/s)     &          90.49  \\
&\\
\hline
\multicolumn{2}{c}{ESO/NTT} \\
\hline
&\\
Date              &     2011 10 16  \\
UT                &       03:40:27  \\
Exptime (s)       &            120  \\
$R$               &    $\sim$2,000  \\
&\\
\hline
\multicolumn{2}{c}{Magellan/MIKE} \\
\hline
&\\
Date              &     2015 06 17  \\
UT                &       08:26:20  \\
Exptime (s)       &           1800  \\
$R$               &   $\sim$35,000  \\
v$_{r}$(km/s)     &          94.91 
\enddata
\end{deluxetable}

\subsection{Medium-resolution Spectroscopy}

The medium-resolution spectrum of \hes\ was obtained as part of an effort to
follow-up CEMP candidates from the HES, as described in
\citet{placco2010,placco2011}.  The observations were carried out in semester
2011B using the EFOSC-2 spectrograph \citep{efosc} mounted on the 3.5m ESO New
technology Telescope.  The setup made use of Grism7 (600~gr~mm$^{\rm{-1}}$),
and a 1$\farcs$0 slit with $1\times1$ on-chip binning, resulting
in a wavelength coverage of 3550-5500\,{\AA}, resolving power of $R\sim 2,000$,
dispersion of 0.96\,{\AA}/pixel (with $\sim2$ pixels per
resolution element), and signal-to-noise ratios S/N$\sim 60$ per pixel at
4300\,{\AA}.  The calibration frames included FeAr exposures (taken following
the science observation), quartz-lamp flatfields, and bias frames.  All
reduction tasks were performed using standard
IRAF\footnote{\href{http://iraf.noao.edu}{http://iraf.noao.edu}.} packages.
Table~\ref{candlist} lists basic information on \hes\, and details of the
spectroscopic observations at medium and high resolution.

\subsection{High-resolution Spectroscopy}

A high-resolution spectrum was obtained during the 2015A semester using the
Magellan Inamori Kyocera Echelle \citep[MIKE;][]{mike} spectrograph mounted on
the 6.5m Magellan-Clay Telescope at Las Campanas Observatory.  The observing
setup included a $0\farcs7$ slit with $2\times2$ on-chip binning, yielding a
resolving power of $R\sim35,000$ (blue spectral range) and $R\sim28,000$ (red
spectral range), with $\sim3$ pixels per resolution element. The
S/N at $4300\,${\AA} is $\sim100$ per pixel.  MIKE spectra have nearly full
optical wavelength coverage ($\sim3500-8500\,${\AA}).  The data were reduced
using the data reduction pipeline developed for MIKE spectra, first described
by
\citet{kelson2003}\footnote{\href{http://code.obs.carnegiescience.edu/python}
{http://code.obs.carnegiescience.edu/python}}.

\vspace{1.0cm}

\section{Analysis}
\label{secatm}

\subsection{Stellar Parameters}

The stellar atmospheric parameters were first obtained from the
medium-resolution ESO/NTT spectrum using the n-SSPP \citep{beers2014}, a
modified version of the SEGUE Stellar Parameter Pipeline
\citep[SSPP;][]{lee2008a,lee2008b,allende2008,lee2011, smolinski2011,lee2013}.
The values for \teff, \logg, and \metal\ determined from this analysis were
used as first estimates for the high-resolution analysis.
Using the high-resolution MIKE spectrum, we determined the stellar parameters
spectroscopically, using software developed by \citet{casey2014}.
Equivalent-width measurements were obtained by fitting Gaussian profiles to the
observed absorption lines. Table~\ref{eqw} lists the lines used in this work,
their measured equivalent widths, and the derived abundance from each line.  We
employed one-dimensional plane-parallel model atmospheres with no overshooting
\citep{castelli2004}, computed under the assumption of local thermodynamic
equilibrium (LTE). 

\begin{deluxetable}{lrrrrr}[!ht]
\tabletypesize{\tiny}
\tablewidth{0pc}
\tablecaption{\label{eqw} Equivalent-Width Measurements}
\tablehead{
\colhead{Ion}&
\colhead{$\lambda$}&
\colhead{$\chi$} &
\colhead{$\log\,gf$}&
\colhead{$W$}&
\colhead{$\log\epsilon$\,(X)}\\
\colhead{}&
\colhead{({\AA})}&
\colhead{(eV)} &
\colhead{}&
\colhead{(m{\AA})}&
\colhead{}}
\startdata
\\
      C CH  & 4246.000 & \nodata & \nodata &  syn &    5.83 \\  %
      C CH  & 4313.000 & \nodata & \nodata &  syn &    5.73 \\  %
      N CN\xx  & 3883.000 & \nodata & \nodata & syn &  6.13 \\  %
      Na I  & 5889.950 & 0.00 &    0.108 &  97.06 &    2.77 \\
      Na I  & 5895.924 & 0.00 & $-$0.194 &  83.35 &    2.79 \\
      Mg I  & 3829.355 & 2.71 & $-$0.208 &    syn &    4.60 \\
      Mg I  & 3832.304 & 2.71 &    0.270 &    syn &    4.60 \\
      Mg I  & 3838.292 & 2.72 &    0.490 &    syn &    4.60 \\
      Mg I  & 4571.096 & 0.00 & $-$5.688 &    syn &    4.90 \\
      Mg I  & 4702.990 & 4.33 & $-$0.380 &    syn &    4.90 \\
\enddata
\tablenotetext{a}{Using \eps{C}=5.78}
\tablecomments{Table available in its entirety in machine-readable form.}
\end{deluxetable}

The effective temperature of \hes\ was determined by minimizing trends between
the abundances of 77 \ion{Fe}{1} lines and their excitation potentials, and
applying the temperature corrections suggested by \citet{frebel2013}.
The microturbulent velocity was determined by minimizing the trend between the
abundances of \ion{Fe}{1} lines and their reduced equivalent widths.  The
surface gravity was determined from the balance of the two ionization stages of
iron, \ion{Fe}{1} and \ion{Fe}{2}. 
\hes\ also had its stellar atmospheric parameters determined from the 
moderate-resolution ($R\sim8,000$) RAVE spectrum by
\citet{kordopatis2013}. These values, together with our determinations
from the medium- and high-resolution spectra, are listed in Table~\ref{obstable}.

There is very good agreement between the temperatures derived from the
medium- and high-reslution spectra used in this work; the RAVE value is
about $\sim 200$~K warmer. The surface gravities are all within 2$\sigma$,
and the ESO/NTT and RAVE values agree with each other exactly.
This is expected, since both of these estimates come from isochrone matching, 
while the high-resolution \logg\ was determined from the balance of \ion{Fe}{1}
and \ion{Fe}{2} lines. For \metal, the RAVE value is 0.8~dex higher than the
other two results. According to the RAVE Data Release 4 \citep{kordopatis2013},
the spectrum for \hes\ has a S/N=40, which is below the S/N=50
limit set by \citet{kordopatis2011} for reliable parameter estimates for halo
giants, and their pipeline analysis converged without warnings for this star.
Since the difference in temperature cannot alone account for such a large
discrepancy in \metal, this most likely arises from a combination
of the lower S/N and the methods used to calibrate the RAVE metallicity
scale for giants, which have a 0.40~dex dispersion in their
residuals.

\begin{deluxetable}{lcccc}[!ht]
\tablewidth{0pc}
\tabletypesize{\scriptsize}
\tablecaption{Derived Stellar Parameters for \protect\hes \label{obstable}}
\tablehead{
\colhead{              }&
\colhead{\teff{}(K)    }&
\colhead{\logg{}(cgs)  }&
\colhead{\metal{}      }&
\colhead{$\xi$(km/s)   }}
\startdata
\\
RAVE     & 4974 (101) & 0.95 (0.35) & $-$3.20 (0.10) & \nodata     \\
ESO/NTT  & 4792 (150) & 0.98 (0.35) & $-$4.06 (0.20) & \nodata     \\
Magellan & 4765 (100) & 1.55 (0.20) & $-$4.05 (0.05) & 1.50 (0.20) \\
\enddata
\end{deluxetable}

\begin{deluxetable}{lrrrrcr}[!ht]
\tablewidth{0pc}
\tabletypesize{\scriptsize}
\tablecaption{Abundances for Individual Species \label{abund}}
\tablehead{
Species  & $\log\epsilon_{\odot}$\,(X) & $\log\epsilon$\,(X) 
         & $\mbox{[X/H]}$              & $\mbox{[X/Fe]}$    
		 & $\sigma$                    & $N$ }
\startdata
\\
C (CH)   &  $+$8.43 &  $+$5.78 &  $-$2.65 &            $+$1.40\hphantom{\yy}    &    0.15 &  2 \\
C (CH)   &  $+$8.43 &  $+$6.12 &  $-$2.31 &            $+$1.74\xx               &    0.15 &  2 \\
N (CN)   &  $+$7.83 &  $+$6.13 &  $-$1.70 &            $+$2.35\hphantom{\yy}    &    0.20 &  1 \\
Na I     &  $+$6.24 &  $+$2.78 &  $-$3.46 &            $+$0.59\hphantom{\yy}    &    0.05 &  2 \\
Mg I     &  $+$7.60 &  $+$4.58 &  $-$3.02 &            $+$1.03\hphantom{\yy}    &    0.10 &  8 \\
Al I     &  $+$6.45 &  $+$2.05 &  $-$4.40 &            $-$0.26\hphantom{\yy}    &    0.10 &  1 \\
Si I     &  $+$7.51 &  $+$4.36 &  $-$3.19 &            $+$0.90\hphantom{\yy}    &    0.10 &  1 \\
Ca I     &  $+$6.34 &  $+$2.70 &  $-$3.64 &            $+$0.40\hphantom{\yy}    &    0.05 &  4 \\
Sc II    &  $+$3.15 &  $-$0.72 &  $-$3.87 &            $+$0.18\hphantom{\yy}    &    0.05 &  3 \\
Ti I     &  $+$4.95 &  $+$1.17 &  $-$3.78 &            $+$0.27\hphantom{\yy}    &    0.05 &  4 \\
Ti II    &  $+$4.95 &  $+$1.19 &  $-$3.76 &            $+$0.28\hphantom{\yy}    &    0.10 & 11 \\
Cr I     &  $+$5.64 &  $+$1.51 &  $-$4.13 &            $-$0.09\hphantom{\yy}    &    0.05 &  5 \\
Mn I     &  $+$5.43 &  $+$1.43 &  $-$4.00 &            $+$0.05\hphantom{\yy}    &    0.05 &  4 \\
Fe I     &  $+$7.50 &  $+$3.45 &  $-$4.05 & \hphantom{$-$}0.00\hphantom{\yy}    &    0.05 & 77 \\
Fe II    &  $+$7.50 &  $+$3.45 &  $-$4.05 & \hphantom{$-$}0.00\hphantom{\yy}    &    0.08 &  3 \\
Co I     &  $+$4.99 &  $+$1.29 &  $-$3.70 &            $+$0.35\hphantom{\yy}    &    0.05 &  4 \\
Ni I     &  $+$6.22 &  $+$1.94 &  $-$4.28 &            $-$0.23\hphantom{\yy}    &    0.05 &  7 \\
Sr II    &  $+$2.87 &  $-$1.91 &  $-$4.78 &            $-$0.73\hphantom{\yy}    &    0.07 &  2 \\
Ba II    &  $+$2.18 &  $-$2.97 &  $-$5.15 &            $-$1.11\hphantom{\yy}    &    0.05 &  2 \\
Eu II    &  $+$0.52 & $<-$3.28 & $<-$3.80 &           $<+$0.25\hphantom{\yy}    & \nodata &  1 \\
Pb I     &  $+$1.75 & $<-$0.25 & $<-$2.00 &           $<+$2.05\hphantom{\yy}    & \nodata &  1 \\
\enddata
\tablenotetext{a}{\cfe=$+$1.74 using corrections of \citet{placco2014c}.}
\end{deluxetable}

\begin{figure*}[!ht]
\epsscale{1.15}
\plotone{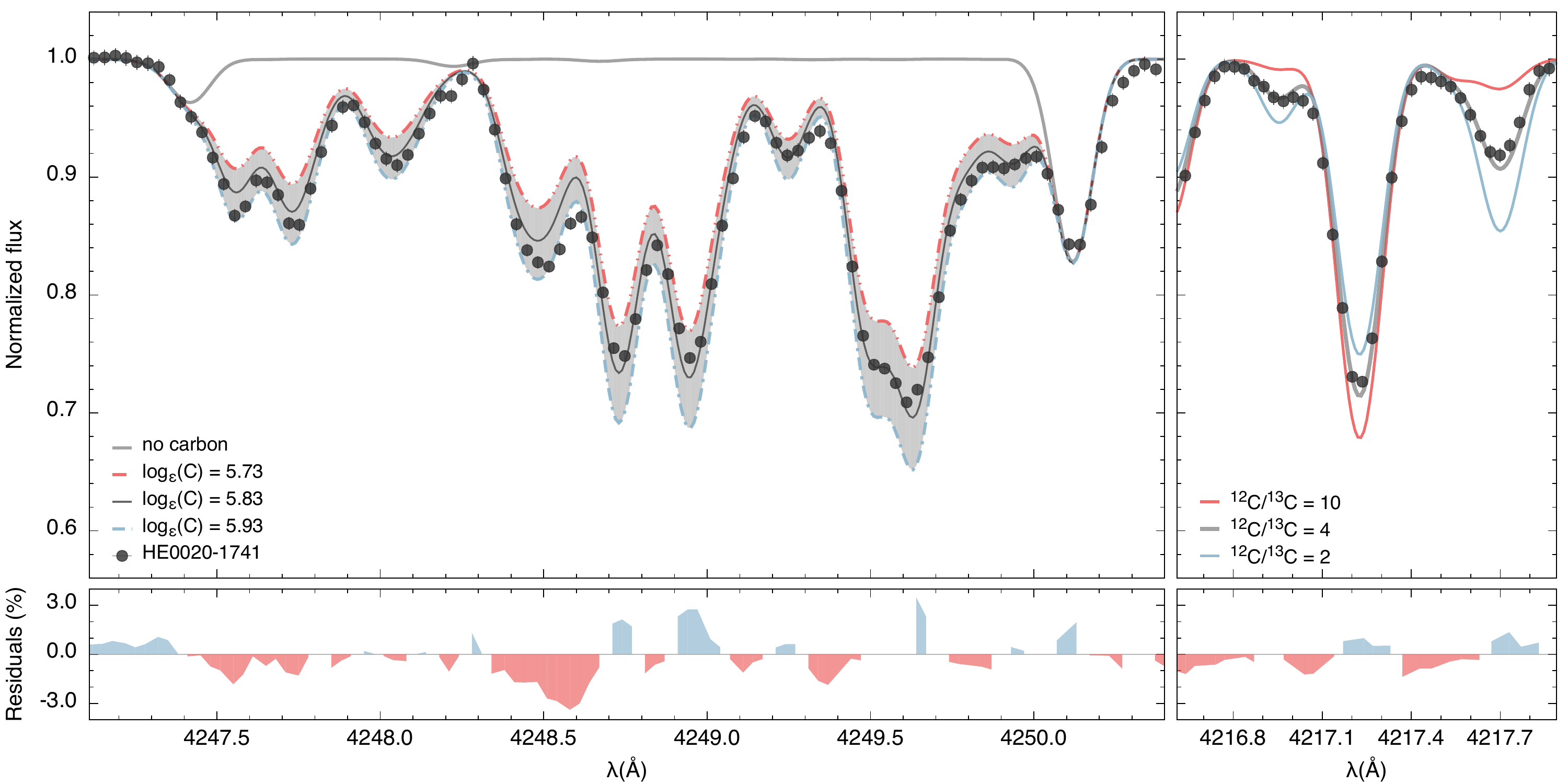}
\caption{
Left panel: Spectral synthesis of the CH $G$-band for \protect\hes.  The dots
represent the observed high-resolution spectrum, the solid line is the best 
abundance fit, and
the dotted and dashed lines are the lower and upper abundances, used to
estimate the abundance uncertainty. The gray shaded area encompasses a 0.2~dex
difference in \eps{C}. The light gray line shows the synthesized spectrum in
the absence of carbon. Right panel: Determination of the carbon isotopic
ratio, $^{12}$C/$^{13}$C. The dots represent the observed spectrum, the solid gray
line is the best fit, and the colored lines are the lower and upper abundances, used to
estimate the uncertainty. The lower panels show the residuals (in \%) between
the observed data and the best abundance fit.}
\label{csyn}
\end{figure*}

\subsection{Chemical Abundances and Upper Limits}

Elemental-abundance ratios, \xfe{X}, are calculated adopting Solar photospheric
abundances from \citet{asplund2009}.  The average measurements (or
upper limits) for $18$ elements, derived from the Magellan/MIKE spectrum, are
listed in Table~\ref{abund}.  The $\sigma$ values are the standard
error of the mean.  Abundances were calculated by both equivalent-width analysis
and spectral synthesis. The 2014 version of the MOOG code \citep{sneden1973},
which includes a more realistic treatment of scattering \citep[see][for further
details]{sobeck2011}, is used for the spectral synthesis.

\subsubsection{Carbon and Nitrogen}
\label{cnsec}

Carbon abundance was derived from CH molecular features at
$\lambda$4246 (\eps{C}=5.83) and $\lambda$4313 (\eps{C}=5.73),
with an average value of \eps{C}=5.78 (\cfe $= +1.40$). 
The left panels of Figure~\ref{csyn} show the spectral synthesis of the
CH $G$-band for \hes. The dots represent the observed spectra, the solid
line is the best abundance fit, and the dotted and dashed lines are the
lower and upper abundances, used to estimate the uncertainty. The gray line
shows the synthesized spectrum in the absence of carbon. The lower panel
shows the residuals (in \%) between the observed data and the best
abundance fit, which are all below 3\% for the synthesized region. %
Since \hes\ is on the upper red-giant branch, the observed carbon
abundance does not reflect the chemical composition of its natal gas
cloud. By using the procedure described in \citet{placco2014}, 
which interpolates the observed \logg, \metal, and \cfe\ of \s01\ 
with a grid of theoretical stellar evolution models for low-mass stars,
we determine the carbon depletion due to CN processing for \hes\ to be
0.34~dex.

The $^{12}$C/$^{13}$C isotopic ratio is a sensitive indicator of the extent of
mixing processes in stars on the red-giant branch.  Using a fixed elemental
carbon abundance (\eps{C}=5.78) for the CH features around $\lambda$4217\,{\AA},
we derived $^{12}$C/$^{13}$C = 4, which suggests that substantial processing of
$^{12}$C into $^{13}$C has taken place in the star.
We note that this ratio also indicates that considerable processing
of carbon into nitrogen has occured in \s01.
The right panels of Figure~\ref{csyn} show the determination of the
$^{12}$C/$^{13}$C isotopic ratio. The dots represent the observed
spectrum, the solid gray line is the best fit, with the two other values 
taken to be lower and upper limits. The lower panel shows that the residuals between
the observed data and $^{12}$C/$^{13}$C = 4 are all below 2\%.

The nitrogen abundance was determined from spectral synthesis of the CN band at
$\lambda$3883\,{\AA}. The NH band at $\lambda$3360\,{\AA} did not
have sufficiently high S/N to allow for a proper spectral synthesis.
For the CN band, we used a fixed carbon 
abundance of \eps{C}=5.78 (average of carbon abundances determined from the CH
band), and derived a value of \eps{N}=6.13, with an uncertainty of $\pm 0.2$~dex.

\subsubsection{From Na to Ni}

Abundances of Na, Sc, Ti, Cr, Co, and Ni were determined by
equivalent-width analysis only. For Ti, where transitions from two
different ionization stages were measured, the abundances agree within
0.02~dex. Spectral synthesis was used to determine abundances for Mg,
Al, Si, Ca, and Mn \citep[accounting for hyperfine splitting from the
linelists from][]{denhartog2011}.

\subsubsection{Neutron-capture Elements}

The chemical abundances for Sr and Ba, as well as upper limits for Eu
and Pb, were determined via spectral synthesis. We used the compilation
of linelists by \citet{frebel2014}, based on lines from
\citet{aoki2002}, \citet{barklem2005}, \citet{lawler2009}, and the VALD 
database \citep{vald}. The neutron-capture absorption lines in the blue
spectral region, particularly close to strong CH or CN features, need to be
carefully synthesized, since these are intrinsically weak in CEMP-no stars.
Because of that, we included the observed carbon and nitrogen
abundances for all the syntheses, as well as the $^{12}$C/$^{13}$C isotope ratio.

The Sr abundance was determined from the $\lambda$4077
(\eps{Sr} = $-$1.93) and $\lambda$4215 (\eps{Sr} = $-$1.88) lines, with an
average value of \xfe{Sr} = $-$0.73. For Ba, both $\lambda$4554 and
$\lambda$4934 features were successfully synthesized with
\eps{Ba} = $-$2.97 (\xfe{Ba} = $-$1.11). Both Sr and Ba abundances are
within typical ranges for CEMP-no stars
\citep{placco2014}. 

Upper limits were determined for Eu ($\lambda$4129) and Pb
($\lambda$4057). The Pb upper limit (\xfe{Pb}$ < +$2.05) is similar to the
one for the CEMP-no \bd\ \citep{placco2014b}, and it adds further
evidence that the origin of the neutron-capture abundances in \hes\ is
unlikely to be from an unseen evolved companion. The \xfe{Pb} should be
higher by at least a factor of ten to agree with theoretical predictions
for the $s$-process \citep{bisterzo2010}. Furthermore, the
radial-velocity monitoring of \hes\ reported by
\citet{hansen2016} revealed no significant variation ($\sigma = 0.212$ km\
s$^{-1}$) over a temporal window of 1066 days.

\begin{deluxetable}{lrrrrr}
\tabletypesize{\scriptsize}
\tablewidth{0pc}
\tablecaption{Example Systematic Abundance Uncertainties for \protect\hes \label{sys}}
\tablehead{
\colhead{Elem}&
\colhead{$\Delta$\teff}&
\colhead{$\Delta$\logg}&
\colhead{$\Delta\xi$}&
\colhead{$\sigma/\sqrt{n}$}&
\colhead{$\sigma_{\rm tot}$}\\
\colhead{}&
\colhead{$+$150\,K}&
\colhead{$+$0.3 dex}&
\colhead{$+$0.3 km/s}&
\colhead{}&
\colhead{}}
\startdata
\\
 Na I   &    0.14   & $-$0.04   & $-$0.11   &    0.07   &    0.20 \\
 Mg I   &    0.13   & $-$0.12   & $-$0.07   &    0.04   &    0.19 \\
 Al I   &    0.11   & $-$0.07   & $-$0.11   &    0.10   &    0.20 \\
 Si I   &    0.14   & $-$0.02   & $-$0.01   &    0.10   &    0.17 \\
 Ca I   &    0.10   & $-$0.02   & $-$0.01   &    0.05   &    0.11 \\
Sc II   &    0.09   &    0.07   & $-$0.04   &    0.06   &    0.13 \\
 Ti I   &    0.17   & $-$0.03   & $-$0.01   &    0.05   &    0.18 \\
Ti II   &    0.06   &    0.06   & $-$0.08   &    0.03   &    0.12 \\
 Cr I   &    0.17   & $-$0.04   & $-$0.06   &    0.04   &    0.19 \\
 Mn I   &    0.19   & $-$0.05   & $-$0.07   &    0.07   &    0.22 \\
 Fe I   &    0.16   & $-$0.05   & $-$0.09   &    0.01   &    0.19 \\
Fe II   &    0.01   &    0.09   & $-$0.02   &    0.08   &    0.12 \\
 Co I   &    0.18   & $-$0.03   & $-$0.02   &    0.05   &    0.19 \\
 Ni I   &    0.17   & $-$0.07   & $-$0.09   &    0.04   &    0.21 \\
Sr II   &    0.09   &    0.06   & $-$0.10   &    0.07   &    0.16 \\
Ba II   &    0.12   &    0.08   & $-$0.01   &    0.07   &    0.16 \\
\enddata
\end{deluxetable}

\subsection{Uncertainties}

Uncertainties in the elemental-abundance determinations, as well as the
systematic uncertainties due to changes in the atmospheric parameters, were
treated in the same way as described in \citet{placco2013,placco2015b}.
Table~\ref{sys} shows how variations within the quoted uncertainties in each
atmospheric parameter affect the derived chemical abundances.  Also listed is
the total uncertainty for each element, which is calculated from the quadratic
sum of the individual error estimates.  Even though \teff, \logg, and $\xi$\
are correlated, we assume complete knowledge of two of them to assess how
changes in the third parameter would affect the abundance calculation. For
example, a change in $+$150~K in \teff\ for \s01\ requires a change of about
0.1~dex in \logg\ to maintain the balance between \ion{Fe}{1} and \ion{Fe}{2}.
However, this change in \logg\ translates to a 0.01~dex change in abundance,
which is 16 times smaller than the change due to \teff. Then, for simplicity,
we assume that the variables are independent, and use the quadratic sum.  For
this calculation, we used spectral features with abundances determined by
equivalent-width analysis only.  The variations for the parameters are $+$150~K
for \teff, $+$0.3~dex for \logg, and $+$0.3 km\,s$^{-1}$ for $\xi$.  

\section{Discussion}
\label{disc}

\subsection{Model Predictions for UMP Progenitors}

In this section, we assess the main properties of the possible stellar
progenitors of selected UMP stars from the literature, by comparing their
abundance patterns with theoretical model predictions for Pop III
stars. We employ the non-rotating massive-star models from
\citet{heger2010}, for which the free parameters are the mass ($10\,
\Msun-100\,\Msun$), explosion energies ($0.3-10\times10^{51}\,\erg$),
and the amount of mixing in the SNe ejecta. The grid used in the
work has 16,800 models, and the $\chi^2$ matching algorithm is described
in \citet{heger2010}. An online tool with the model database can be
accessed at
\texttt{starfit}\footnote{\href{http://starfit.org}{http://starfit.org}}.
For the present application, we adopt a similar procedure as the one
described in \citet{placco2015} and \citet{roederer2016}, as
described below.

\begin{figure*}[!ht]
\epsscale{1.15}
\plottwo{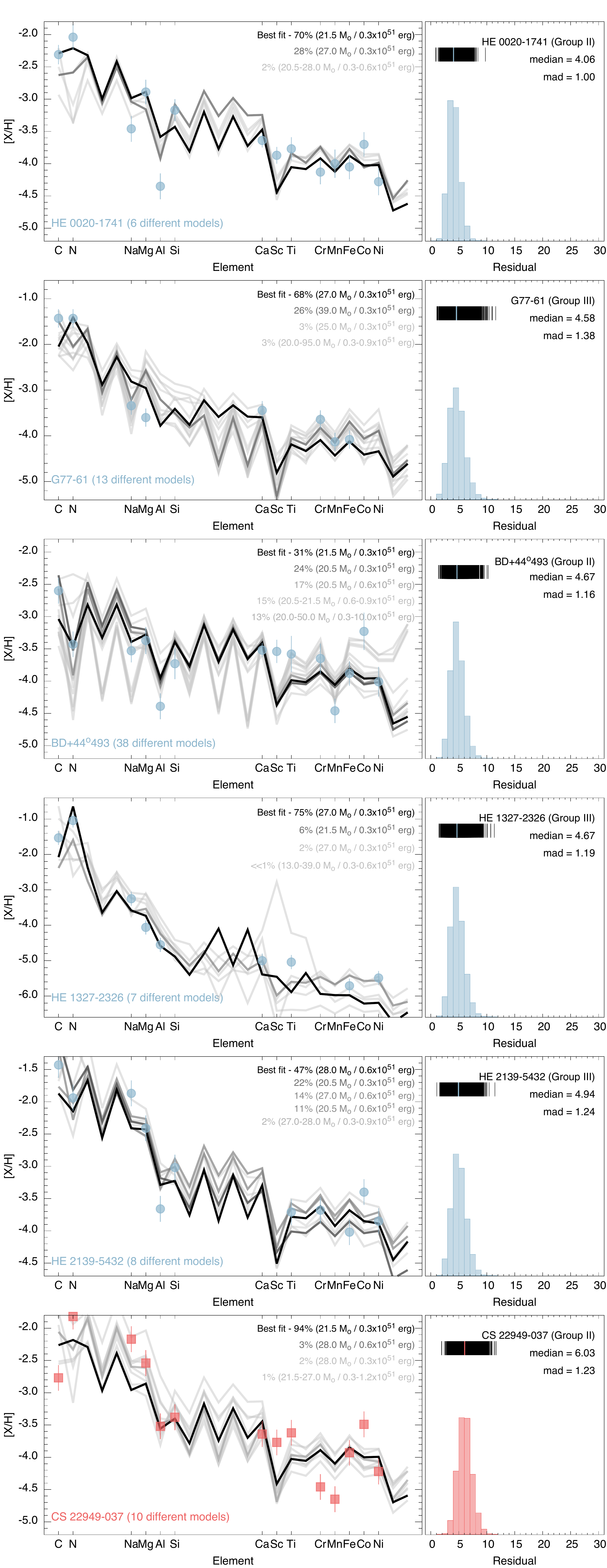}{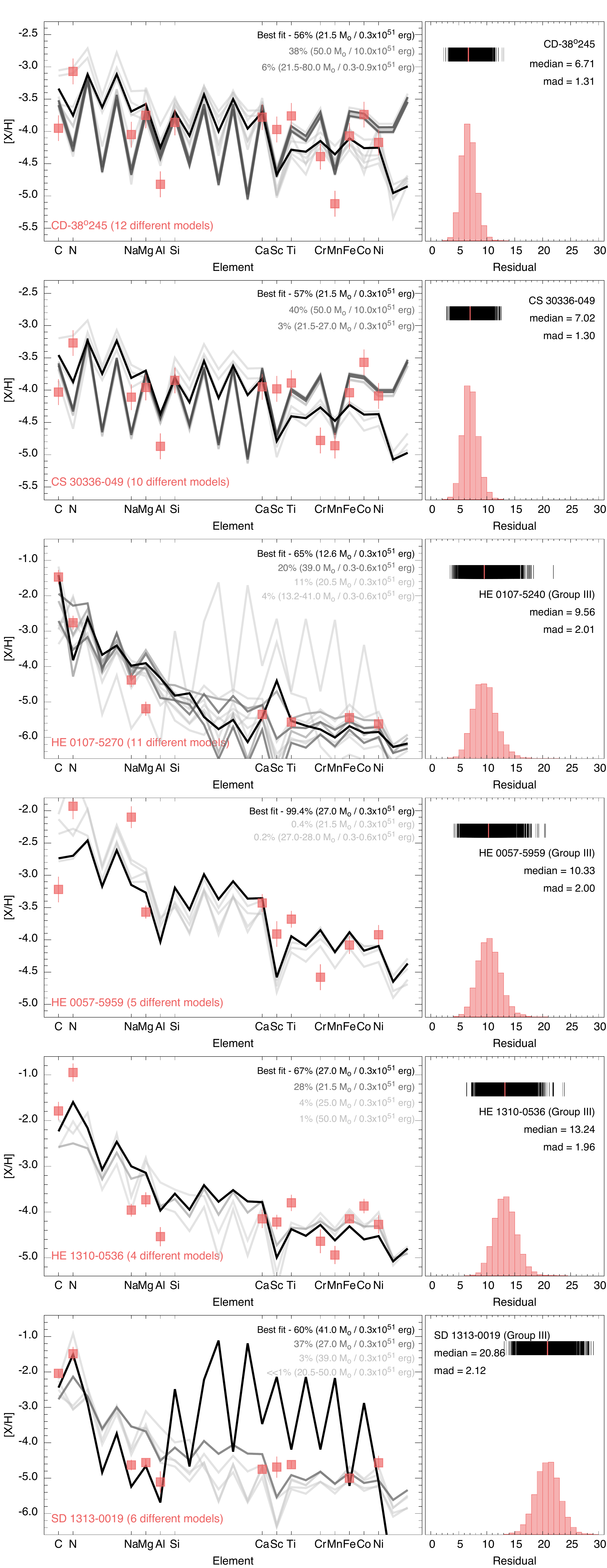}
\caption{
Best model fits for \protect\hes\ and the 11 other UMP stars from the literature, ordered
by increasing median residuals. For each star, the left panel shows the
simulated abundance patterns and models, where the masses and explosion energies
are provided in the legend at the upper right of each panel, color-coded
by their fractional occurence. The right panel shows the distribution of the residuals for
the 10,000 simulations for each star. The colored bar overlaying the upper
density distributions in the right panels marks the median value, which is
shown in the legend of each panel at the top right, along with the $mad$ (median absolute deviation).} 
\label{group1}
\end{figure*}

\begin{figure*}[!ht]
\epsscale{1.15}
\plotone{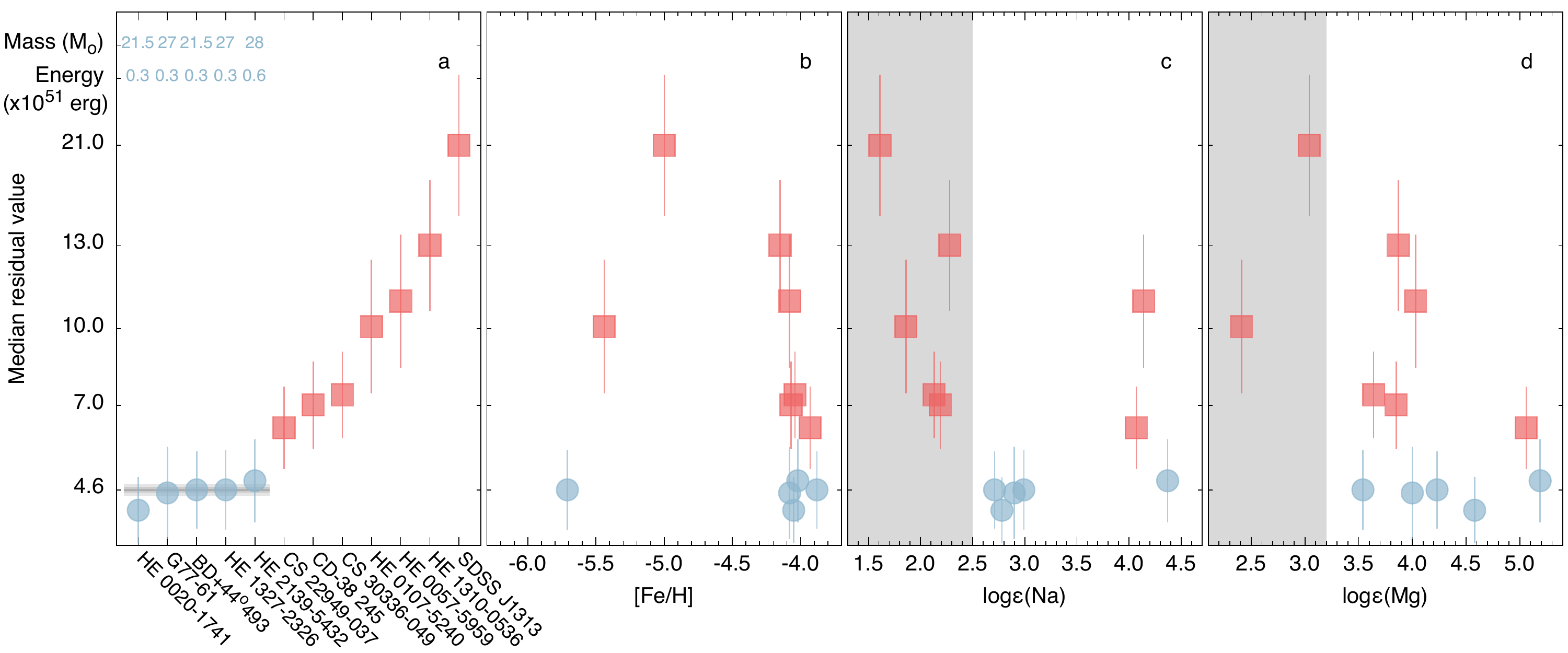}
\caption{
Median residuals for \protect\hes\ and 11 other UMP stars from the literature,
color-coded by the division proposed in Figure~\ref{group1}.  The stars are
ordered by increasing median values (Panel~$a$).  The solid horizontal line
shows the median for the five stars with low residuals, and the gray shaded
areas encompass $\pm1\times mad$ and $\pm2\times mad$.  The numbers on the top
are the progenitor masses (in M$_{\odot}$) and explosion energy
($\times10^{51}\,\erg$) from the best model fit for each star. Panels~$b$, $c$,
and $d$ show the residuals as a function of \metal, \eps{Na}, and
\eps{Mg}, respectively.
} 
\label{residuals}
\end{figure*}

We collected UMP stars from the literature for which at least carbon, nitrogen,
sodium, magnesium, calcium, and iron were measured. With \metal$ < -4.0$, only 11
stars have abundances for these elements determined from high-resolution
spectroscopy (including \hes). We also added \bd\ to the sample,
since its metallicity is \metal\ = $-$3.8, and it has a number of abundances
measured with small uncertainties. Table~\ref{litdata} lists the abundances and
references for the literature sample. The stars are divided according to the
CEMP-no Groups II and III from \citet{yoon2016}; two of our sample stars
are classified as carbon-normal stars. The carbon
abundances were corrected for the C depleted by CN processing, following the
procedures described in \citet{placco2014c}, and briefly summarized
in Section~\ref{cnsec}. In addition, we calculated the
inverse corrections for the nitrogen abundances, by imposing that the
[(C$+$N)/Fe] ratio remains constant throughout the evolution of the star on the
giant branch. The model grid we employed is the same as that in \citet{placco2014c}.
Even though there may also be ON processing that occurs in the
star, the effect is expected to be negligible compared to the CN processing
\citep[see bottom panel of Figure 2 in][]{placco2014c}.

For the model matching, we have used the following abundances (where
available): C, N, Na, Mg, Al, Si, Ca, Sc, Ti, V, Cr, Mn, Fe, Co, and Ni.
For consistency, no upper limits were used. For each star, we assembled $10^4$
unique abundance patterns, based on the observed values from the literature.
For each element, we generated $10^4$ random numbers from a normal distribution, 
with the measured abundances as the central value, and the uncertainties as the
dispersion. 
Then, we randomly selected individual abundances from each
distribution, and generated new abundance patterns.  For each of these, we
found the progenitor mass, explosion energy, and the mean squared residuals
from the \texttt{starfit} code. 

Figure~\ref{group1} shows the results of this exercise for the 12 stars
in Table~\ref{litdata}. For each star, the left panel shows the best
model fits for the $10^4$ abundance patterns. The masses and explosion
energies of the models are given in the legend at the upper right in
each panel, color-coded by their fractional occurence. The right-hand
panel for each star shows the distribution of the mean squared residuals
for the simulations. The colored bars overlaying the upper density
distributions on the right panels mark the median residual value, and
its value is shown in the legend of these panel at the top right, along
with the $mad$ (median absolute deviation), a robust estimator of the
dispersion. The stars are ordered by increasing median values.

As an example, consider the top left panel of Figure~\ref{group1}, for the 
star \s01.  We first ran the {\texttt{starfit}} code to find the best model 
fit for each of the $10^4$ abundance patterns generated from the observed 
abundances (blue filled circles). In about 70\% (7,041) of the resampled 
abundance patterns, the best fit was the model with $21.5\, \Msun$ and 
$0.3\times10^{51}\,\erg$ (black solid line). In 2,849 (28\%) of the patterns,
the model with $27.0\, \Msun$ and $0.3\times10^{51}\,\erg$ gave the best fit
(dark gray solid line), and for the remaining 110 abundance patterns (light gray
solid lines), the best-fit models had masses between $20.5-28.0\, \Msun$ and 
explosion energies between $0.3-0.6\times10^{51}\,\erg$. In total, 6 unique
models (out of the 16,800 models on the grid) were able to account for the best
fits for the $10^4$ resampled abundance patterns. For each best fit, the
{\texttt{starfit}} code calculates the mean squared residual, and the
distribution of these values is given on the panel to the right side of the
abundance patterns. Above the distribution is the density plot (black stripes),
with the median value (blue stripe) shown on the upper right part of the panel,
together with the $mad$.

For the first five stars in the left-hand column of panels shown in
Figure~\ref{group1}, the mass and explosion energy of the progenitor are within
$20.5-28.0\, \Msun$ and $0.3-0.6\times10^{51}\,\erg$. These values are
consistent with predictions from \citet{nomoto2006} for the faint supernova
progenitor scenario. Considering the two most-frequent models for each star,
the C-N abundances are well-reproduced in most cases, within $2\sigma$ of the
theoretical values. The Na-Si abundances agree within $2\sigma$ with the best
model fits (except for Al in \s01\ and Mg in \s02). The Ca and Ti abundances
also agree within $2\sigma$, except for Ti in \s04. The large over-abundances
of Sc for \s01\ and \s03\ are expected, since Sc may have contributions from
other nucleosynthesis processes \citep[see][for further details]{heger2010}.
Among the Cr-Fe abundances, there is also overall good agreement, with Fe
values agreeing within $1\sigma$ with the best model fits. The enhanced cobalt
abundances for \s01, \s03, and \s05\ cannot be accounted for by the models, but
we note that \eps{Co} is often under-predicted by theoretical models
\citep[e.g.,][]{tominaga2014,roederer2016}.

For stars in the right-hand panels of Figure~\ref{group1} (red filled squares),
the progenitor masses range from $12.6\,\Msun$ for \s09\ to $41.0\,\Msun$ for
\s12, all with explosions energies of $0.3\times10^{51}\,\erg$. For these
stars, as reflected by their residual distributions, there is a clear mismatch
between the model predictions and the observed abundances. The stars \s06,
\s07, and \s08\ all have best fits with a $21.5\, \Msun$ and
$0.3\times10^{51}\,\erg$ model. This is likely a numerical artifact, since the
C and N abundances for these three stars are in complete disagreement with the
models. For \s10, only the Ca abundance matches the best model yields, and most
values are at least 0.5~dex away from the theoretical values. For \s11, the
only observed abundance within $1\sigma$ of the model is Fe. In this case, it
is clear that no models can reproduce the large difference between the C-N and
Na-Al abundances. For \s09\ and \s12, the models are not able to simultaneously
predict the C-N abundances and the Na-Mg or Ca-Ti abundances.

Based on these results, we infer that the first five stars shown in the
left-hand column of panels of Figure~\ref{group1} have abundance patterns that
are consistent with the progenitor stellar population described by the models
of \citet{heger2010}. For the remaining seven stars, the poor agreement between
observations and models suggests the presence of at least one additional class
of stellar progenitors for UMP stars.

Figure~\ref{residuals} shows the behavior of the median residual values for
\hes\ and the data for other UMP stars from the literature, where the stars are
ordered by increasing median values (Panel~$a$), and as a function of \metal\
(Panel~$b$), \eps{Na}\ (Panel~$c$), and \eps{Mg} (Panel~$d$).
In Panel~$a$,
the gray shaded areas encompass $\pm1\times mad$ and $\pm2\times mad$ for the
five stars (blue filled circles) in the left-hand column of
Figure~\ref{group1}.
The numbers shown in the legend are the progenitor masses (in M$_{\odot}$) and
the explosion energy ($\times10^{51}\,\erg$) from the best model fit for each
star. The gray regions highlighted on Panels~$c$ and $d$ are explained below.

Distinctions between the stars with low (blue symbols) and high (red symbols)
residuals become more evident when inspecting the median residuals in
Figure~\ref{residuals}. Even though this separation does not appear to be
correlated with \metal\ (Panel~$b$), observations of additional stars are
needed to cover the gap between $-5.0 \leq $~\metal~$ \leq -4.0$, in order to
uncover any  possible trend.  Nevertheless, \s04\ (\metal~=~$-5.71$) has a
median residual value consistent with that of other stars at \metal~$\sim-4.0$,
and it does not agree with the values for \s09\ (\metal~=~$-5.44$) and \s12\
(\metal~=~$-5.00$).  From inspection of Panel~$c$, it can be seen that the
low-residual stars exhibit \eps{Na}~$\gtrsim2.5$, and are mostly concentrated
at \eps{Na}~$\sim2.8$. The only exception is \s05, with \eps{Na}~$ = 4.37$.  
The trend for Na can be an indication that
another stellar progenitor, producing (but not limited to)
\eps{Na}~$\lesssim2.5$ (gray shaded area on
Panel~$c$), is needed to account for the diferences in the observed
abundances.  
In Panel~$d$, even though the low-residual stars exhibit \eps{Mg}~$\gtrsim3.5$,
there is no obvious separation between these and the high-residual stars. The
two high-residual stars with low \eps{Na} in Panel~$c$ also have low \eps{Mg}.
However, this correlation does not hold for the other stars in the same group.

\begin{figure*}[!ht]
\epsscale{1.15}
\plotone{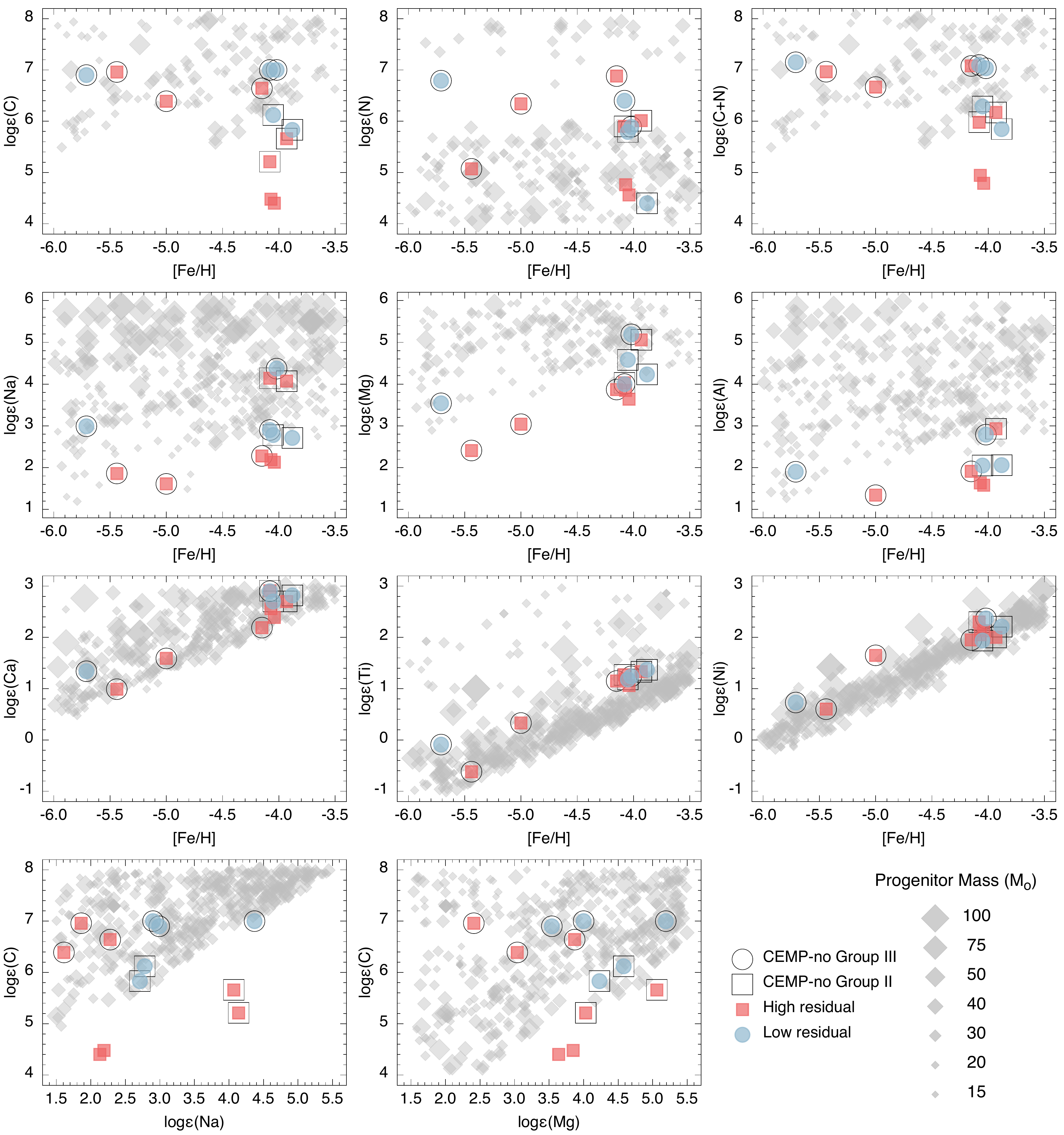}
\caption{Comparison between the chemical abundances of the literature sample of
UMP stars (colored symbols) and the predicted yields from the supernovae models
used in this work (gray symbols). The progenitor mass is proportional to the
size of the symbol. The open symbols relate to the CEMP-no classifications
proposed by \citet{yoon2016}.}
\label{abunda}
\end{figure*}

\begin{deluxetable*}{lcccccccccccccccc}
\tabletypesize{\scriptsize}
\tablewidth{0pc}
\tablecaption{Literature data (\protect\eps{X} abundances)\label{litdata}}
\tablehead{
\colhead{Star name}&
\colhead{C}& \colhead{N}&
\colhead{Na}&\colhead{Mg}&
\colhead{Al}&\colhead{Si}&
\colhead{Ca}&\colhead{Sc}&
\colhead{Ti}&\colhead{Cr}&
\colhead{Mn}&\colhead{Fe}&
\colhead{Co}&\colhead{Ni}&
\colhead{Ref.}&
\colhead{Note}}
\startdata
\multicolumn{17}{c}{CEMP-no Group II}\\
\hline
BD$+$44$^\circ$493 & 5.83 & 4.40 & 2.71 & 4.23 &    2.06 &    3.78 &    2.82 & $-$0.39 & \hphantom{$-$}1.37 &    1.99 &    0.97 & 3.62 &    1.76 &    2.21 &  1 & \nodata \\
CS~22949$-$037     & 5.66 & 6.01 & 4.07 & 5.06 &    2.93 &    4.13 &    2.70 & $-$0.62 & \hphantom{$-$}1.33 &    1.18 &    0.78 & 3.57 &    1.50 &    2.00 &  2 &       a \\
HE~0020$-$1741     & 6.12 & 5.79 & 2.78 & 4.58 &    2.05 &    4.36 &    2.70 & $-$0.72 & \hphantom{$-$}1.18 &    1.51 &    1.43 & 3.45 &    1.29 &    1.94 &  3 &       b \\
HE~0057$-$5959     & 5.21 & 5.90 & 4.14 & 4.03 & \nodata & \nodata &    2.91 & $-$0.76 & \hphantom{$-$}1.27 &    1.06 & \nodata & 3.42 & \nodata &    2.30 &  4 & \nodata \\
\hline
\multicolumn{17}{c}{CEMP-no Group III}\\
\hline
G77$-$61           & 7.00 & 6.40 & 2.90 & 4.00 & \nodata & \nodata &    2.90 & \nodata &            \nodata &    2.00 &    1.30 & 3.42 & \nodata & \nodata &  5 & \nodata \\
HE~0107$-$5240     & 6.96 & 5.57 & 1.86 & 2.41 & \nodata & \nodata &    0.99 & \nodata &            $-$0.62 & \nodata & \nodata & 2.06 & \nodata &    0.60 &  6 &       c \\
HE~1310$-$0536     & 6.64 & 6.88 & 2.28 & 3.87 &    1.91 & \nodata &    2.19 & $-$1.07 & \hphantom{$-$}1.15 &    1.00 &    0.49 & 3.35 &    1.12 &    1.95 &  7 &       d \\
HE~1327$-$2326     & 6.90 & 6.79 & 2.99 & 3.54 &    1.90 & \nodata &    1.34 & \nodata &            $-$0.09 & \nodata & \nodata & 1.79 & \nodata &    0.73 &  8 & \nodata \\
HE~2139$-$5432     & 7.00 & 5.89 & 4.37 & 5.19 &    2.79 &    4.49 & \nodata & \nodata & \hphantom{$-$}1.24 &    1.96 & \nodata & 3.48 &    1.59 &    2.37 &  9 & \nodata \\
SDSS~J1313$-$0019  & 6.39 & 6.34 & 1.61 & 3.04 &    1.34 & \nodata &    1.59 & $-$1.54 & \hphantom{$-$}0.33 & \nodata & \nodata & 2.50 & \nodata &    1.65 & 10 & \nodata \\
\hline
\multicolumn{17}{c}{Carbon-Normal Stars}\\
\hline
CD$-$38$^\circ$245 & 4.48 & 4.76 & 2.19 & 3.85 &    1.63 &    3.65 &    2.56 & $-$0.82 & \hphantom{$-$}1.19 &    1.25 &    0.31 & 3.43 &    1.25 &    2.05 & 11 &       e \\
CS~30336$-$049     & 4.40 & 4.56 & 2.13 & 3.64 &    1.58 &    3.66 &    2.39 & $-$0.83 & \hphantom{$-$}1.06 &    0.86 &    0.57 & 3.46 &    1.42 &    2.13 & 12 &       f
\enddata
\tablerefs{\\
(1):  \citet{ito2013}; \\
(2):  \citet{cohen2013}; \\
(3):  This work; \\
(4):  \citet{yong2013}; \\
(5):  \citet{plez2005}; \\
(6):  \citet{christlieb2004}; \\
(7):  \citet{hansen2015}; \\
(8):  \citet{frebel2008}; \\
(9):  \citet{yong2013}; \\
(10): \citet{frebel2015b}. \\
(11): \citet{francois2003}; \\
(12): \citet{lai2008}; \\
}
\tablecomments{\\
(a):  $\Delta$\eps{C} = $+0.15$, $\Delta$\eps{N} = $-0.09$; \\
(b):  $\Delta$\eps{C} = $+0.34$, $\Delta$\eps{N} = $-0.34$; \\
(c):  Average of C and N values; \\
(d):  $\Delta$\eps{C} = $+0.07$, $\Delta$\eps{N} = $0.00$; \\
(e):  \eps{C} from \citet{mcw1995}, \eps{N} from \citet{spite2006}, 
      $\Delta$\eps{C} = $+0.07$, $\Delta$\eps{N} = $-0.04$; \\
(f):  $\Delta$\eps{C} = $+0.29$, $\Delta$\eps{N} = $-0.15$; \\
}
\end{deluxetable*}

\subsection{Abundance Comparison between UMP Stars and Yields from Massive
Metal-free Stars}

To further investigate the differences among the twelve UMP stars, we compared
the individual abundances as a function of \metal, \eps{Na}, and \eps{Mg}\ with
the yields from the 16,800 models used for the matching procedure.
Figure~\ref{abunda} shows the result of this exercise. The colored symbols are
the observed abundances from the literature sample (including \hes), and the
gray symbols are yields from the supernova models. The open symbols show the
CEMP-no groups proposed in Paper I. For the theoretical values, the
progenitor mass is proportional to the size of the symbol. Models with abundance
values outside the ranges shown in Figure~\ref{abunda} were suppressed for
simplicity.

Overall, the twelve stars have abundances consistent with the models for the
elements Ca, Ti, and Ni. This also holds true for other elements not shown in
Figure~\ref{abunda}, such as Cr, Mn, and Co.
Abundances for the five low-residual stars (blue filled circles) present, in
general, good agreement with the values predicted by models with M $ <
30\,\Msun$. Some exceptions include: \eps{N} for \s02\ and \s04\ (however,
\eps{C+N} agrees well in both cases); \eps{Al} for \s01\ and \s03\ is about
0.5~dex lower than the model values. Even though the Al abundance was determined
from spectral synthesis, that region ($\lambda$3961\,{\AA}) presents strong CH
and CN absorption features, which could compromise the determination. 
For the seven stars with higher median residual values, there are clear
mismatches between the observations and the theoretical predictions. For \s07\
and \s08, carbon and nitrogen are at least 1~dex lower than the model values,
and Na, Mg, and Al are also below the model ranges. Even though \s06\ exhibits
better agreement for C and N, the two bottom panels show that four stars are
below the model expectations for \eps{C} vs.  \eps{Na}\ and \eps{C} vs. \eps{Mg}.
There is also no agreement for Na, Mg, and Al (\s09\ and \s12) as a function of
metallicity. 

Inspection of the two bottom panels of Figure~\ref{abunda} reveals that C, Na,
and Mg abundances 
may suggest some deficiencies in the current models, which could 
exclude these as possible progenitors for four of the seven high-residual
stars. In both cases, the carbon abundances are over-predicted by the models
when compared to the observations. This suggests that the current models do not
adequately capture the progenitors of the full set of UMP stars.
The combination of carbon, sodium, and magnesium was used in Paper~I as part of
the justification for the division of the CEMP-no stars in Groups~II and III.
When comparing the low- and high-residual stars with their CEMP-no group
classifications, it can be seen that half of the Group~II stars (\s09\ and
\s12) are outside the model ranges for \eps{C} vs. \eps{Na} and \eps{C} vs.
\eps{Mg}. The two other Group~II stars (\s01\ and \s03) have low median
residuals, and are close to the limits explored by the models. 
Another interesting case is \s04\ (Group~III), which is the most metal-poor
star in this sample; it presents a low median residual, unlike the other stars
with \metal$ < -5.0$. Its chemical abundances are in regions where the model
grid is sparse, specially for C, N, Na, and Mg. Regardless, the models used in
this work cannot account for the abundance patterns of \s09\ and \s12, which
are also Group~III stars based on Paper~I. This could be further evidence that
at least one additional, different progenitor population operates at the lowest
metallicities, such as the spinstars described by \citet{meynet2006}.

For completeness, we also tested the robustness of our fitting results with
respect to the recent study of Ezzeddine et al. (2016, in preparation). This
study suggests the presence of large positive corrections to the Fe abundances
of the 18 most iron-poor stars, following line formation computations in a non
local thermodynamic equilibrium (NLTE). While it would be inconsistent to mix
LTE and NLTE abundances in a given stellar pattern, we tested how trial
corrections for \eps{Fe} of 0.75\,dex to \s04, 0.70\,dex to
\s09, and 0.3\,dex to \s01\ would affect our conclusions. As a result, no
significant changes in the progenitor mass, explosion energy, and mean squared
residuals were found. To properly interpret these new NLTE Fe abundances, we
thus have to await the full NLTE abundance patterns and compare those with the
supernova yields, once available.

\vspace{1.0cm}

\section{Conclusions}
\label{final}

We have presented the first high-resolution spectroscopic study of the ultra
metal-poor, CEMP-no star \hes. This star adds to the small number of stars with
\metal\ $ < -4.0$ identified to date, and presents the same behavior as most
(more than 80\%) of UMP stars: \cfe$ > +1.0$ and \xfe{Ba}$<0.0$.
We have attempted to characterize the progenitor stellar population of UMP
stars by comparing the yields from a grid of metal-free, massive-star models
with the observed elemental abundances for \hes\ and 11 other UMP stars from
the literature for which at least abundances of C, N, Na, Mg, Al, and Fe have
been reported. 

Based on our residual analysis, we find that 42\% (5 of 12) of the sample stars
have elemental-abundance patterns that are well-reproduced by the SNe models of
\citet{heger2010}.  We conclude that this class of SNe explosions and their
associated nucleosynthesis cannot alone account for the observed abundance
patterns of the entire set of UMP stars we have considered. In particular,
Pop III, massive metal-free star models cannot reproduce abundance
patterns of stars with \eps{C} $ \leq 5.0$, \eps{Na}~ $ \lesssim 2.5$, and
\eps{Mg}~ $ \lesssim 3.2$.  Carbon and sodium could potentially be
used as tracers of the stellar progenitor, in addition to the metallicity. This
could be evidence of different progenitor signatures, such as fast-rotating
massive Pop III stars from \citet{meynet2006}, or the faint SNe from
\citet{nomoto2006}. If we assume that the models used in this work are a viable
candidate for the stellar progenitors of UMP stars, there must exist {\it{at
least}} one additional class of progenitor operating at \metal$<-4.0$, or a
single process, yet to be identified, that could account for the abundances of
all UMP stars. We should also acknowledge the possibility that more than one
progenitor could contribute to the observed abundance patterns of the
high-residual stars presented in this work.

Recent evidence presented by \citet{yoon2016} suggests that the carbon abundance
may be key to differentiate between different stellar progenitors of UMP stars.
The classifications based on absolute carbon abundances, $A$(C), are further
confirmed when looking at the sodium and magnesium abundances. The analysis
presented in this paper supports this hypothesis, and suggests that the
progenitor population(s) for UMP stars may be even more rich and complex than
previously thought.
We emphasize that abundances for additional UMP stars, in particular those in
the metallicity regime $-5.0 < $ [Fe/H] $ < -4.0$, are needed to better
constrain the nature of the possible stellar progenitors. Nitrogen abundances
are an important constraint to the theoretical models, and currently more than
half of the observed UMP stars have only limits for this element reported.

\acknowledgments

We thank Chris Sneden, who kindly provided part of the linelists used for the
spectral synthesis, Nozomu Tominaga, for clarifying specific characteristics
of the faint SNe models, Richard Stancliffe, for clarifications on the stellar
evolution models used for the CN corrections, and the anonymous referee, 
who made valuable
suggestions that helped improve the paper.
V.M.P., T.C.B., and J.Y. \ acknowledge partial support for this work from the
National Science Foundation under Grant No. PHY-1430152 (JINA Center for the
Evolution of the Elements).
A.F. and A.C.\ are supported by NSF CAREER grant AST-1255160.  
A.H.\ was supported by a Future Fellowship of the Australian Research Council
(FT120100363).
This research has made use of NASA's Astrophysics Data System Bibliographic
Services; the arXiv pre-print server operated by Cornell University; the SIMBAD
database hosted by the Strasbourg Astronomical Data Center; the IRAF software
packages distributed by the National Optical Astronomy Observatories, which are
operated by AURA, under cooperative agreement with the NSF; the SAGA Database
\citep[Stellar Abundances for Galactic Archeology;][]{saga2008}; the
{\texttt{R-project}} software package \citep{rproject}; the {\texttt{gnuplot}}
command-line plotting program \citep{gnuplot}; and the online Q\&A platform
{\texttt{stackoverflow}} (\href{http://stackoverflow.com/}{http://stackoverflow.com/}).

\bibliographystyle{apj}

\clearpage
\ltab

\begin{deluxetable}{lrrrrr}
\tabletypesize{\tiny}
\tablewidth{0pc}
\tablecaption{\label{eqwfull} Equivalent-Width Measurements}
\tablehead{
\colhead{Ion}&
\colhead{$\lambda$}&
\colhead{$\chi$} &
\colhead{$\log\,gf$}&
\colhead{$W$}&
\colhead{$\log\epsilon$\,(X)}\\
\colhead{}&
\colhead{({\AA})}&
\colhead{(eV)} &
\colhead{}&
\colhead{(m{\AA})}&
\colhead{}}
\startdata
      C CH  & 4246.000 & \nodata & \nodata &  syn &    5.83 \\  %
      C CH  & 4313.000 & \nodata & \nodata &  syn &    5.73 \\  %
      N CN\xx  & 3883.000 & \nodata & \nodata & syn &  6.13 \\  %
      Na I  & 5889.950 & 0.00 &    0.108 &  97.06 &    2.77 \\
      Na I  & 5895.924 & 0.00 & $-$0.194 &  83.35 &    2.79 \\
      Mg I  & 3829.355 & 2.71 & $-$0.208 &    syn &    4.60 \\
      Mg I  & 3832.304 & 2.71 &    0.270 &    syn &    4.60 \\
      Mg I  & 3838.292 & 2.72 &    0.490 &    syn &    4.60 \\
      Mg I  & 4571.096 & 0.00 & $-$5.688 &    syn &    4.90 \\
      Mg I  & 4702.990 & 4.33 & $-$0.380 &    syn &    4.90 \\
      Mg I  & 5172.684 & 2.71 & $-$0.450 &    syn &    4.25 \\
      Mg I  & 5183.604 & 2.72 & $-$0.239 &    syn &    4.25 \\
      Mg I  & 5528.405 & 4.34 & $-$0.498 &    syn &    4.85 \\
      Al I  & 3961.520 & 0.01 & $-$0.340 &    syn &    2.05 \\ %
      Si I  & 4102.936 & 1.91 & $-$3.140 &    syn &    4.36 \\ %
      Ca I  & 4454.780 & 1.90 &    0.260 &    syn &    2.64 \\ %
      Ca I  & 5588.760 & 2.52 &    0.210 &   4.87 &    2.71 \\
      Ca I  & 6122.220 & 1.89 & $-$0.315 &   8.78 &    2.75 \\
      Ca I  & 6162.170 & 1.90 & $-$0.089 &  11.92 &    2.68 \\
     Sc II  & 4246.820 & 0.32 &    0.240 &  57.45 & $-$0.75 \\
     Sc II  & 4400.389 & 0.61 & $-$0.540 &  13.70 & $-$0.69 \\
     Sc II  & 4415.544 & 0.59 & $-$0.670 &  10.63 & $-$0.72 \\
      Ti I  & 3989.760 & 0.02 & $-$0.062 &  10.92 &    1.14 \\
      Ti I  & 4533.249 & 0.85 &    0.532 &   6.36 &    1.19 \\
      Ti I  & 4981.730 & 0.84 &    0.560 &   7.55 &    1.17 \\
      Ti I  & 4991.070 & 0.84 &    0.436 &   6.31 &    1.20 \\
     Ti II  & 3759.291 & 0.61 &    0.280 &  85.34 &    1.10 \\
     Ti II  & 3761.320 & 0.57 &    0.180 &  84.26 &    1.12 \\
     Ti II  & 4012.396 & 0.57 & $-$1.750 &  26.82 &    1.30 \\
     Ti II  & 4417.714 & 1.17 & $-$1.190 &  23.80 &    1.31 \\
     Ti II  & 4443.801 & 1.08 & $-$0.720 &  39.21 &    1.07 \\
     Ti II  & 4450.482 & 1.08 & $-$1.520 &  17.08 &    1.34 \\
     Ti II  & 4468.517 & 1.13 & $-$0.600 &  42.03 &    1.07 \\
     Ti II  & 4501.270 & 1.12 & $-$0.770 &  37.73 &    1.13 \\
     Ti II  & 4533.960 & 1.24 & $-$0.530 &  42.09 &    1.12 \\
     Ti II  & 4563.770 & 1.22 & $-$0.960 &  29.43 &    1.25 \\
     Ti II  & 4571.971 & 1.57 & $-$0.320 &  38.54 &    1.23 \\
      Cr I  & 4254.332 & 0.00 & $-$0.114 &  57.03 &    1.50 \\
      Cr I  & 4274.800 & 0.00 & $-$0.220 &  52.38 &    1.48 \\
      Cr I  & 4289.720 & 0.00 & $-$0.370 &  49.12 &    1.56 \\
      Cr I  & 5206.040 & 0.94 &    0.020 &  19.37 &    1.48 \\
      Cr I  & 5208.419 & 0.94 &    0.160 &  26.73 &    1.52 \\
      Mn I  & 4041.380 & 2.11 & $-$0.350 &    syn &    1.43 \\ %
      Mn I  & 4754.021 & 2.28 & $-$0.647 &    syn &    1.43 \\ %
      Mn I  & 4783.424 & 2.30 & $-$0.736 &    syn &    1.43 \\ %
      Mn I  & 4823.514 & 2.32 & $-$0.466 &    syn &    1.43 \\ %
      Fe I  & 3727.619 & 0.96 & $-$0.609 &  70.55 &    3.28 \\
      Fe I  & 3743.362 & 0.99 & $-$0.790 &  65.97 &    3.32 \\
      Fe I  & 3753.611 & 2.18 & $-$0.890 &  16.00 &    3.41 \\
      Fe I  & 3758.233 & 0.96 & $-$0.005 &  97.02 &    3.50 \\
      Fe I  & 3763.789 & 0.99 & $-$0.221 &  80.31 &    3.23 \\
      Fe I  & 3765.539 & 3.24 &    0.482 &  13.97 &    3.20 \\
      Fe I  & 3767.192 & 1.01 & $-$0.390 &  78.19 &    3.35 \\
      Fe I  & 3786.677 & 1.01 & $-$2.185 &  15.44 &    3.30 \\
      Fe I  & 3820.425 & 0.86 &    0.157 & 105.31 &    3.33 \\
      Fe I  & 3824.444 & 0.00 & $-$1.360 &  82.76 &    3.28 \\
      Fe I  & 3886.282 & 0.05 & $-$1.080 &  96.13 &    3.48 \\
      Fe I  & 3887.048 & 0.91 & $-$1.140 &  57.68 &    3.22 \\
      Fe I  & 3899.707 & 0.09 & $-$1.515 &  80.32 &    3.39 \\
      Fe I  & 3902.946 & 1.56 & $-$0.442 &  57.99 &    3.30 \\
      Fe I  & 3917.181 & 0.99 & $-$2.155 &  25.67 &    3.51 \\
      Fe I  & 3920.258 & 0.12 & $-$1.734 &  72.13 &    3.33 \\
      Fe I  & 3922.912 & 0.05 & $-$1.626 &  83.66 &    3.57 \\
      Fe I  & 3940.878 & 0.96 & $-$2.600 &   9.80 &    3.39 \\
      Fe I  & 3949.953 & 2.18 & $-$1.251 &   9.97 &    3.49 \\
      Fe I  & 3977.741 & 2.20 & $-$1.120 &  14.87 &    3.58 \\
      Fe I  & 4005.242 & 1.56 & $-$0.583 &  57.44 &    3.38 \\
      Fe I  & 4045.812 & 1.49 &    0.284 &  91.47 &    3.49 \\
      Fe I  & 4063.594 & 1.56 &    0.062 &  80.91 &    3.45 \\
      Fe I  & 4071.738 & 1.61 & $-$0.008 &  71.90 &    3.28 \\
      Fe I  & 4132.058 & 1.61 & $-$0.675 &  56.36 &    3.47 \\
      Fe I  & 4143.414 & 3.05 & $-$0.200 &   8.23 &    3.32 \\
      Fe I  & 4143.868 & 1.56 & $-$0.511 &  64.30 &    3.46 \\
      Fe I  & 4147.669 & 1.48 & $-$2.071 &  12.23 &    3.55 \\
      Fe I  & 4181.755 & 2.83 & $-$0.371 &  14.24 &    3.51 \\
      Fe I  & 4187.039 & 2.45 & $-$0.514 &  20.15 &    3.40 \\
      Fe I  & 4187.795 & 2.42 & $-$0.510 &  25.89 &    3.51 \\
      Fe I  & 4191.430 & 2.47 & $-$0.666 &  13.35 &    3.35 \\
      Fe I  & 4202.029 & 1.49 & $-$0.689 &  60.77 &    3.43 \\
      Fe I  & 4216.184 & 0.00 & $-$3.357 &  31.11 &    3.60 \\
      Fe I  & 4227.427 & 3.33 &    0.266 &  11.13 &    3.31 \\
      Fe I  & 4233.603 & 2.48 & $-$0.579 &  18.59 &    3.44 \\
      Fe I  & 4250.787 & 1.56 & $-$0.713 &  65.76 &    3.67 \\
      Fe I  & 4260.474 & 2.40 &    0.077 &  42.71 &    3.27 \\
      Fe I  & 4337.046 & 1.56 & $-$1.695 &  20.27 &    3.51 \\
      Fe I  & 4375.930 & 0.00 & $-$3.005 &  36.62 &    3.33 \\
      Fe I  & 4383.545 & 1.48 &    0.200 &  93.96 &    3.46 \\
      Fe I  & 4404.750 & 1.56 & $-$0.147 &  84.25 &    3.60 \\
      Fe I  & 4415.122 & 1.61 & $-$0.621 &  62.06 &    3.48 \\
      Fe I  & 4427.310 & 0.05 & $-$2.924 &  50.84 &    3.61 \\
      Fe I  & 4447.717 & 2.22 & $-$1.339 &  10.82 &    3.58 \\
      Fe I  & 4459.118 & 2.18 & $-$1.279 &  14.36 &    3.62 \\
      Fe I  & 4461.653 & 0.09 & $-$3.194 &  31.59 &    3.50 \\
      Fe I  & 4494.563 & 2.20 & $-$1.143 &  14.62 &    3.51 \\
      Fe I  & 4528.614 & 2.18 & $-$0.822 &  23.98 &    3.44 \\
      Fe I  & 4531.148 & 1.48 & $-$2.101 &  14.00 &    3.59 \\
      Fe I  & 4602.941 & 1.49 & $-$2.208 &  13.52 &    3.68 \\
      Fe I  & 4871.318 & 2.87 & $-$0.362 &  13.81 &    3.44 \\
      Fe I  & 4872.137 & 2.88 & $-$0.567 &   8.68 &    3.43 \\
      Fe I  & 4890.755 & 2.88 & $-$0.394 &  11.25 &    3.37 \\
      Fe I  & 4891.492 & 2.85 & $-$0.111 &  19.51 &    3.35 \\
      Fe I  & 4918.994 & 2.85 & $-$0.342 &  13.04 &    3.36 \\
      Fe I  & 4920.503 & 2.83 &    0.068 &  24.34 &    3.28 \\
      Fe I  & 5012.068 & 0.86 & $-$2.642 &  22.92 &    3.60 \\
      Fe I  & 5041.756 & 1.49 & $-$2.200 &   9.41 &    3.44 \\
      Fe I  & 5051.634 & 0.92 & $-$2.764 &  16.69 &    3.61 \\
      Fe I  & 5083.339 & 0.96 & $-$2.842 &  12.69 &    3.59 \\
      Fe I  & 5142.929 & 0.96 & $-$3.080 &   9.26 &    3.67 \\
      Fe I  & 5171.596 & 1.49 & $-$1.721 &  24.59 &    3.46 \\
      Fe I  & 5194.942 & 1.56 & $-$2.021 &  14.64 &    3.55 \\
      Fe I  & 5216.274 & 1.61 & $-$2.082 &  10.02 &    3.48 \\
      Fe I  & 5232.940 & 2.94 & $-$0.057 &  17.98 &    3.33 \\
      Fe I  & 5269.537 & 0.86 & $-$1.333 &  80.63 &    3.60 \\
      Fe I  & 5328.039 & 0.92 & $-$1.466 &  74.91 &    3.64 \\
      Fe I  & 5328.531 & 1.56 & $-$1.850 &  21.81 &    3.59 \\
      Fe I  & 5371.489 & 0.96 & $-$1.644 &  65.45 &    3.61 \\
      Fe I  & 5397.128 & 0.92 & $-$1.982 &  44.81 &    3.44 \\
      Fe I  & 5405.775 & 0.99 & $-$1.852 &  45.96 &    3.42 \\
      Fe I  & 5429.696 & 0.96 & $-$1.881 &  45.71 &    3.40 \\
      Fe I  & 5434.524 & 1.01 & $-$2.126 &  37.72 &    3.56 \\
      Fe I  & 5446.917 & 0.99 & $-$1.910 &  45.31 &    3.46 \\
      Fe I  & 5455.609 & 1.01 & $-$2.090 &  38.32 &    3.53 \\
      Fe I  & 5615.644 & 3.33 &    0.050 &  10.91 &    3.39 \\
     Fe II  & 4583.840 & 2.81 & $-$1.930 &  15.01 &    3.63 \\
     Fe II  & 4923.930 & 2.89 & $-$1.320 &  25.42 &    3.38 \\
     Fe II  & 5018.450 & 2.89 & $-$1.220 &  28.24 &    3.34 \\
      Co I  & 3845.468 & 0.92 &    0.010 &  33.28 &    1.33 \\
      Co I  & 3995.306 & 0.92 & $-$0.220 &  21.90 &    1.24 \\
      Co I  & 4118.767 & 1.05 & $-$0.490 &  12.50 &    1.33 \\
      Co I  & 4121.318 & 0.92 & $-$0.320 &  21.33 &    1.29 \\
      Ni I  & 3500.850 & 0.17 & $-$1.294 &  38.99 &    1.92 \\
      Ni I  & 3519.770 & 0.28 & $-$1.422 &  32.90 &    2.00 \\
      Ni I  & 3566.370 & 0.42 & $-$0.251 &  61.60 &    1.90 \\
      Ni I  & 3597.710 & 0.21 & $-$1.115 &  44.05 &    1.88 \\
      Ni I  & 3783.520 & 0.42 & $-$1.420 &  32.18 &    2.01 \\
      Ni I  & 3807.140 & 0.42 & $-$1.220 &  36.58 &    1.91 \\
      Ni I  & 3858.301 & 0.42 & $-$0.951 &  51.19 &    1.99 \\
     Sr II  & 4077.714 & 0.00 &    0.150 &    syn & $-$1.93 \\ %
     Sr II  & 4215.524 & 0.00 & $-$0.180 &    syn & $-$1.88 \\ %
     Ba II  & 4554.033 & 0.00 &    0.163 &    syn & $-$2.97 \\ %
     Ba II  & 4934.086 & 0.00 & $-$0.160 &    syn & $-$2.97 \\ %
     Eu II  & 4129.720 & 0.00 &    0.220 &    syn & $<-$3.28 \\ %
     Pb  I  & 4057.810 & 1.32 & $-$0.170 &    syn & $<-$0.25 \\ %
\enddata
\tablenotetext{a}{Using \eps{C} = 5.78}
\end{deluxetable}

\end{document}